\documentclass[12pt,preprint]{aastex}
\def\lsim{\mathrel{\rlap{\lower4pt\hbox{\hskip1pt$\sim$}}
    \raise1pt\hbox{$<$}}}                
\def\gsim{\mathrel{\rlap{\lower4pt\hbox{\hskip1pt$\sim$}}
    \raise1pt\hbox{$>$}}}                

\begin{document}


\title{VLBA Observations of G5.89-0.39: OH masers and magnetic 
field structure}


\author {D.P. Stark\altaffilmark{1,3}, 
W.M. Goss\altaffilmark{2}, 
E. Churchwell \altaffilmark{3},
V.L. Fish \altaffilmark{2,5},
I. M. Hoffman 
\altaffilmark{4}
} 

\altaffiltext{1}{Department of Astronomy, California Institute of Technology, MS 105-24, Pasadena, CA 91125; dps@astro.caltech.edu}
\altaffiltext{2}{National Radio Astronomy Observatory, P.O. Box 0, 1003 
Lopezville Road, Socorro, NM 87801}
\altaffiltext{3}{Astronomy Department, University of Wisconsin,
475 N. Charter St., Madison, WI 53706}
\altaffiltext{4}{St. Paul's School, 325 Pleasant Street, Concord, NH 03301}
\altaffiltext{5}{Jansky Fellow}

\begin{abstract}

We present VLBA observations of 1667 MHz OH maser emission from the massive 
star formation region G5.89-0.39.  The observations were phase referenced 
allowing the absolute positions of the masers to be obtained.  
The 1667 MHz masers have 
radial velocities that span $\sim$50 km s$^{-1}$ but show little evidence of 
tracing the bipolar molecular outflow, as has been claimed in 
previous studies.  
We identify 23 Zeeman pairs through comparison of masers 
in left and right circular polarization.  Magnetic field 
strengths range from $-$2 mG to +2 mG, and an ordered reversal 
in magnetic field 
direction is observed toward the southern region of the UC HII region.  
We suggest that the velocity and magnetic field structure of 
the 1667 MHz masers can be explained in the context of a model in 
which the masers arise in a neutral shell just outside a rapidly 
exanding ionized shell.  
 
\end{abstract}

\keywords{HII regions -- ISM: individual (G5.89-0.39) magnetic fields -- masers -- 
stars: formation}

\section{Introduction}

Massive stars are born in optically thick, dense molecular clouds.  Since 
optical emission from young embedded massive stars is completely absorbed 
by surrounding
gas and dust, other means must be used to identify the formation process
of massive stars; hence, massive star formation regions are often 
identified by 
one or more of the following: ultra-compact regions of ionized hydrogen 
(UC HII regions), powerful molecular outflows, and strong molecular masers. 
Studies of maser emission toward star formation regions offer several 
advantages. The high intensity and small spatial scales of maser spots 
allow star formation regions to be probed on scales  $\gsim$ 10$^{14}$ 
cm, providing an effective method of studying the kinematics of gas 
associated with forming stars. Furthermore, Zeeman splitting of maser 
lines provides information about the magnetic field structure of molecular 
gas surrounding young massive stars, offering the possibility to clarify 
the role of magnetic fields in the collimation of bipolar outflows of 
massive stars.

In many of the regions for which hydroxyl (hereafter OH) masers have been 
studied in detail, their velocity structure and distribution suggests that 
they trace a dense disk or molecular torus \citep{Hu99,Hu02,Hu03,Hu05}. 
The magnetic field structure in these regions 
typically shows a toroidal component reversing directions on 
either side of the disk which appears to support the model in which
magnetic field lines are twisted by the rotating disk \citep{U85}. 

A case where the main-line OH masers are predicted to instead trace 
the bipolar outflow is toward 
the massive star formation region G5.89-0.39 (GAL 005.886-00.393 in 
SIMBAD and hereafter referred to as G5.89).  
G5.89 is a shell-like UC HII region, powered by 
a young massive star of spectral type O6 or O7 \citep{WC89}. 
A dense (n$_{H_2}$$\sim$10$^4$ cm$^{-3})$, hot (T$\sim$90 K), and massive 
(M$\sim$30 M$_{\odot}$) envelope of dust and molecular gas surrounds the 
ionized gas \citep{Gomez91,H94} and appears to be 
participating in a powerful bipolar molecular outflow 
\citep{HF88,Ces91,Z90, Ac97,S04,Wat04}.  Proper motion
observations of the expansion velocity of G5.89 suggest a distance of 
$\simeq$2 kpc \citep{Ac98}.  More recently, the distance to 
G5.89 has been estimated kinematically by \cite{Fish03} using HI absorption 
features to resolve the near/far kinematic ambiguity.  The inferred 
distance of $\simeq$3.8 kpc is consistent with the \cite{Ac98} 
estimate within the quoted error bars.  Nevertheless, we adopt 
the \cite{Ac98} distance estimate because the kinematic 
distance method has large uncertainties for sources located near a 
galactic longitude of $\simeq$0$^\circ$. Observations of radio 
recombination lines suggest an LSR velocity of $\simeq$ 10
km s$^{-1}$ \citep{R70,WC89} which we assume to be the systemic 
velocity of G5.89. 

OH maser emission was detected toward G5.89 with the Very Large
Array (VLA) by \cite{Z90} at 1612 MHz, 1665 MHz, and 1667 
MHz. Main-line OH maser spectra in other sources typically show 
many components spanning 
a range of $\sim$10 km s$^{-1}$ in radial velocity; toward G5.89, 
OH maser emission spans a range $\simeq$50 km s$^{-1}$.  The 1667 MHz maser 
features extend primarily N-S, perpendicular to a disk-like 
structure inferred from the morphology of the 6 cm continuum emission. The 
radial velocities of the masers increase with distance away
from the disk-like structure.  \citep{Z90} suggest 
that the main-line OH masers observed 
toward G5.89 trace an accelerating component of the 
bipolar molecular outflow. If this interpretation is valid, 
identification of Zeeman pairs
could yield the magnetic field structure along the outflow,
allowing valuable constraints to be placed on hydromagnetic 
models for bipolar outflows \citep{Shu94}. 

Recent observations indicate that such an interpretation may 
require closer scrutiny.  The outflow position angle differs
considerably depending on the tracer, ranging from NE-SW in 
SiO(5-4) \citep{S04}, to N-S
as traced by C$^{34}$S \citep{Ces91} to E-W (and  
primarily along the line-of-sight) as traced by CO(1-0) and HCO$^+$      
\citep{Wat04}. The source of the outflow has also been debated. 
\cite{Feldt03} detect an O5 star slightly offset from 
the center of the UC HII region and suggest that this object 
may be the ionizing source of G5.89.  \cite{S04} 
argue that the outflow is likely not powered by the O5 star
since the star is not equidistant from the outflow lobes, but rather
originates from the 1.3 mm continuum source detected in their 
work. However, they cannot distinguish whether
the 1.3 mm continuum emission is associated with the UC HII region.  
Studying the maser emission with the spatial resolution offered
by very long baseline interferometry could help clarify these 
uncertainties.  

Since the work of \cite{Z90}, there have been several 
new studies of main-line OH maser emission toward G5.89. \cite{W93}
and \cite{Ar2000} observed 1665 MHz and 1667 MHz OH masers with 
the VLA in left circular polarization (LCP) and right circular 
polarization (RCP). The angular resolution of the \cite{W93} 
and \cite{Ar2000} observations was $\simeq$2\arcsec\  
compared to the  6\arcsec\ resolution
achieved by \cite{Z90}; however, even with 2\arcsec\ resolution, it is 
not possible to reliably identify Zeeman pairs. \cite{Fish05} 
identify Zeeman pairs toward G5.89 
using 1665 MHz and 1667 MHz OH masers with the 
VLBA, but their observations only cover a fraction of the 
velocity-range in which OH masers are known to reside.   

In an attempt to 1) explore the velocity structure of the gas traced by 
the OH masers and 2) characterize the magnetic field of molecular gas 
surrounding G5.89, we have observed the 1667 MHz and 1612 MHz OH maser 
lines toward G5.89 with the Very Long Baseline Array (VLBA) of the 
NRAO$\footnote[1]{The National Radio Astronomy Observatory (NRAO) is a facility
of the National Science Foundation operated under cooperative agreement by 
Associated Universities, Inc.}$. The angular resolution ($\sim$10 milliarcsecs) 
and spectral resolution ($\simeq$0.2 km s$^{-1}$ at 1612 and 1667 MHz) are 
sufficient to 
identify the location of OH maser clumps and identify Zeeman pairs.
The observations and reduction techniques are presented in $\S 2$. In $\S 3$ 
and $\S 4$, we present and discuss the results of our OH maser observations 
toward G5.89. Finally, in $\S 5$, the results are summarized.

\section{Observations}

The 1667.35903 MHz (hereafter 1667 MHz) and 1612.23101 MHz (hereafter 1612 MHz)
 transitions were observed toward G5.89 with the VLBA 
on 2003 May 13-14. Left and right circular polarizations were observed 
simultaneously in 1024 spectral channels covering a total bandwidth of 
1.0 MHz resulting in a channel width of 0.176 km s$^{-1}$ at 1667 MHz and 
0.182 km s$^{-1}$ at 1612 MHz.  
At 1667 MHz, only those channels with emission (500-700) were imaged.  
The resulting velocity resolution 
 (1.2 times the channel separation) was 0.218 km s$^{-1}$ at 1667 MHz and
0.211 km s$^{-1}$ at 1612 MHz. 

Calibrations and imaging were performed with the NRAO AIPS$\footnote[2]{
The Astronomical Image Processing System (AIPS) is documented at http
://www.nrao.edu/aips}$ software package.  
Amplitudes were calibrated using on-line system noise temperature and 
gain curves. Delay and bandpass response solutions were found via 
three scans of the compact continuum source 3C 345. 
The maser observations were phase referenced to J1751-2524 (offset
1.3 degrees from G5.89) using a cycle time of 6 minutes 
(3 minutes on G5.89, 3 minutes on J1751-2524). This source has a 
flux density of 0.67 Jy in the S-band and a positional error of 2.5 mas.
The uncertainty in the absolute position of bright ($\simeq$1 Jy) maser 
spots is thus $\sim$3 mas.  

A single bright maser spot in each maser 
transition and polarization was selected for self-calibration
in each maser transition and polarization. 
Interstellar scatter broadening causes the self-calibration sources 
to become resolved on the longer baselines of the VLBA, significantly 
reducing the signal to noise and causing the calibration to fail 
on these baselines.  The maser used for self-calibration at 1612 MHz 
was brighter than that at 1667 MHz allowing three more antennas to be 
calibrated at 1612 MHz.  The long baselines available from the three extra 
antennas result in a significantly smaller synthesized beam at 1612 MHz.  
The resulting synthesized beams of the 1612 MHz and 1667 MHz images 
are 20 $\times$ 7 mas at a position angle of 7$^\circ$ and 45 $\times$ 
15 mas with a 27$^\circ$ position angle, respectively. At the 
distance of G5.89, this corresponds to $\simeq$30 AU at 1612 MHz and 
$\simeq$50 AU at 1667 MHz.  
The rms noise of the cleaned image is 129 mJy beam$^{-1}$ at 1612 MHz 
and 49 mJy beam$^{-1}$ at 1667 MHz. The observations cover a field 
8.2 $\times$ 8.2 arcsec$^2$ centered on $\alpha_{J2000}$=18$^{h}$
00$^{m}$30\fs382 $\delta_{J2000}$=$-$24$^\circ$04'00\farcs829.  Details of
the observations and instrumental characteristics are presented in 
Table \ref{obs-table1}.  

Using the task JMFIT in AIPS, two-dimensional Gaussians were fit to 
all peaks detected in adjacent channels with significance 
above 5$\sigma$ to identify and 
calculate the flux density of each maser spot.  Line profiles were fit
using IDL software routines.  Maser positions, flux 
densities, and velocities are listed in Table 2 (1667 MHZ LCP), 
Table 3 (1667 MHZ RCP), and Table 4 (1612 MHz).  The maser 
velocities listed in these tables are not corrected for the shift 
induced by the Zeeman effect.

\section{Results}

\subsection{Maser Properties, Position, and Velocity Structure}

Fifty-nine maser spots were identified in LCP and fifty-five in RCP at 
1667 MHz with velocities ranging from $-$32.13 km s$^{-1}$ to +15.30 km s$^{-1}$. 
Figures 1-2 display the positions of the LCP 
and RCP masers, respectively, relative to the VLA 8.5 GHz radio continuum
image (Churchwell 2003, private communication) .  
The offset between the masers and radio continuum
emission is only known at the level of $\simeq$0\farcs3 \citep{Ar2000}, 
however the absolute positions of the masers are known at the 
level of a few mas since they are phase referenced.  
For ease of discussion, we subdivide the OH maser features into three 
groups projected 
on 1) the S-edge of the UC HII (hereafter, G5.89 South) 
and 2) the E-edge of the UC HII region (G5.89 East) 
and 3) the center of the UC HII region (G5.89 Center).  We zoom in
on the distribution of masers in the center of G5.89 in Figure 3
and in the south of G5.89 in Figure 4.  In Figures 
5, 6, and 7 we plot spectra of 1667 MHz averaged over G5.89 South, 
G5.89 East, and G5.89 Center, respectively.  The velocities of maser
features on the edges of the UC HII region are positive and close
to the systemic velocity of 10 km s$^{-1}$, whereas the
velocities of masers projected on the center of the UC HII region are
blueshifted by up to 40 km s$^{-1}$ with respect to the systemic velocity.
In $\S$ 4, we will explain this observation as an optical depth effect.

At 1612 MHz, we observed three masers in both LCP and RCP.
 The positions of 1612 MHz OH masers are presented in Table 4,  
Figure 8 (LCP), and Figure 9 (RCP).  The masers are projected on the 
central region of the 
UC HII region and their line-of-sight velocities are $\simeq-$20 km 
s$^{-1}$. \cite{Z90} identify 1612 MHz OH masers with 
roughly the same projected position and line-of-sight velocity.  They 
also identify several weaker masers at $\simeq$10 km s$^{-1}$ projected on 
the southern region of the UC HII region which are outside our 
velocity coverage and hence 
not detected in our observations.  
The $-$20 km s$^{-1}$ component was first observed by \cite{T69}. 
As with \cite{Z90}, we find that this line is divided in 
several components.  While the ratio of flux densities between the different
components appears to have varied in the fifteen years between 
the \cite{Z90} VLA observations from 1988 and the 2003 VLBA observations 
that make up this work, quantifying the variability is complicated by 
the different resolutions of the VLA and VLBA.

The median deconvolved major axis of the 1667 MHz masers is 16 mas 
in LCP and 17 mas in RCP; both values are roughly consistent with the 
previous measurement of 20 mas in \cite{F06}. At 1612 MHz, the 
median deconvolved major axis is 20 mas in LCP and 21 mas in RCP.  This
suggests that many of the 1667 MHz and 1612 MHZ masers may be 
partially resolved.  However, we note that the deconvolved 
spot parameters may overestimate the true spot size \citep{F06}, 
in part due to scatter broadening.    

The OH maser observations presented in this paper were conducted 
two years after the observations of \cite{Fish05}.  Given the 
large radial velocities of the masers relative to the assumed 
systemic velocity of the system, it is worthwhile to examine whether
there is a positional displacement between the maser spots observed 
at 1667 MHz by \cite{Fish05} and those presented in this paper.
The data presented in \cite{Fish05} were not phase referenced 
so the absolute positions of the masers are unknown. However,
the relative positions are known to very high precision, allowing 
the relative proper motions of masers to be studied.  Since 
\cite{Fish05} only observe positive velocity features, we are limited 
to a subset of the masers presented in this paper.  We identified 
twelve LCP and eleven RCP masers in common between the two datasets. 
The relative proper motions are derived by comparing the relative 
displacement between these maser features in \cite{Fish05} to those 
in this paper; if the masers tangential velocities are large, then 
we would expect the relative displacements to differ between the 
two epochs.  The relative angular displacement between masers in 
G5.89-E and G5.89-S increases by 5-6 mas between the two epochs. 
At the distance of G5.89, 
this proper motion corresponds to 20-30 km s$^{-1}$.  There is 
also significant relative proper motions ($\gsim$4 mas) between 
the individual masers in G5.89 S indicating that these masers 
may be moving at speeds of $\gsim$20 km s$^{-1}$ with respect 
to one another.  

\subsection{Magnetic Field Structure}

A catalog of Zeeman pairs was created by identifying all LCP and RCP lines
that are separated on the sky by less than $\simeq$90 AU, comparable to 
the size of our 
synthesized beam at the distance of G5.89.  Observations of W3(OH) 
\citep{Reid80} suggest that this distance requirement is sufficiently 
rigorous
to reliably derive the magnetic field.   We only classify masers as Zeeman 
pairs if the velocity centroiding error of the pair is less than twice the 
velocity offset between polarizations, 
where we have taken the velocity centroiding error to be the quadrature 
sum of the fitted half-width half maximum divided by twice the 
signal to noise ratio of the lines.    
The magnetic field strength is directly related to 
the velocity offset between the LCP and RCP components: 
B = ($\Delta$v/0.354 km s$^{-1}$) mG at 1667 MHz and B = 
($\Delta$v/0.120 km s$^{-1}$) mG at 1612 MHz.  A positive 
line-of-sight magnetic field (i.e. 
oriented away from the Earth) is assigned to those Zeeman pairs with 
v$_{RCP}$ $>$ v$_{LCP}$, and a negative magnetic field is 
assigned to those with v$_{LCP}$ $>$ v$_{RCP}$.  

Following the method of \cite{Hu03}, we estimate
a $\lsim$ 1.4, 2.0, 6.2 \% probability that any of the Zeeman pairs identified
in G5.89 East, Center, and South are due to random spatial overlap of 
components of opposite polarization.  
 The probability is found by dividing the number of possible pairs 
in a given region (using the same selection criteria as described above)
by the total area of that region and then 
multiplying by the maximum area of separation over which two masers of opposite
polarization are identified as Zeeman pairs (the synthesized beam area). 
The reasonably low probability suggests that the Zeeman pairs in this sample 
are not likely to be false detections. 

We find 23 Zeeman pairs in our sample of 1667 MHz masers. Properties
of the pairs are summarized in Table 5.  All maser velocities 
quoted in the
text and figures of this section are corrected for the 
Zeeman effect.  Spectra of two Zeeman pairs from our observations 
are displayed in Figure 10. 
The velocity offset between the LCP and RCP spectra is clearly visible
in both spectra. In the bottom panel, the flux density 
of the LCP component is lower than that in RCP.  This effect is 
commonly observed in other massive star formation regions \citep{M78}
 and perhaps is a result of velocity or
magnetic field gradients causing differential 
gain for masers of opposite polarizations \citep{Cook68}. In a more 
extreme case, the differential amplification 
between polarizations may result in a maser line being above 
the detection threshold in only one polarization, 
perhaps explaining why just over 
half of the masers in our survey have identifiable Zeeman pairs.  

In Figure 11, we overlay the positions of the Zeeman pairs on the 8.5 
GHz continuum image. Negative line-of-sight magnetic field 
strengths are denoted with circles while positive line-of-sight magnetic 
field strenghs are denoted with crosses. Magnetic field strengths vary from
 $-$1.97 mG to +1.94 mG across the field of G5.89.  

The polarity of the field is positive for all three pairs detected in 
the E group.  Two of the pairs are part of a clump of negative velocity 
features ($-$14.96 and $-$16.43 km s $^{-1}$) with projected
separation of only 75 mas, implying typical linear separations of 
$\sim$150 AU.  The corresponding line-of-sight magnetic field 
strengths are 0.52 mG and 1.34 mG. 
The third Zeeman pair in the east group is offset 770 mas E of the 
other pairs.  The line-of-sight velocity of the pair, 10.03 km s$^{-1}$ is 
significantly offset from velocities of the other two pairs in the region,
while the line-of-sight magnetic field strength, 1.41 mG, is comparable in 
strength and sign to the other pairs.     

Eleven Zeeman pairs are found projected on the central region 
with velocities ranging between $-$31.32 km s$^{-1}$ and 
$-$18.74 km s$^{-1}$.
The radial velocities of most of the pairs in this region are 
centered closely around the median velocity of $-$25.45 km s$^{-1}$.  
The magnetic field polarity 
is positive for all but two of the Zeeman pairs, with amplitudes 
ranging from 0.28 mG to 1.49 mG.  

The field structure derived from the eleven Zeeman pairs in the S
group is more complex than the two previously discussed regions.  
The two southernmost Zeeman pairs have radial velocities ranging from 
13.01 km s$^{-1}$ to 13.91 km s$^{-1}$ and line-of-sight magnetic fields
of negative polarity. Also in the S group is a collection of nine Zeeman 
pairs which display an ordered 
reversal in magnetic field polarity over 300 mas, 
roughly $\sim$600 AU at the distance of G5.89.  Such field reversals may
arise if OH masers trace a rotating molecular torus 
\citep{Hu99,Hu02,Hu03,Hu05}. 
In this scenario, 
the field lines are thought to become entrained in the torus producing the 
observed toroidal component.  This does not appear to be the case toward G5.89
because the the velocity structure does not vary in an 
ordered fashion across the gas traced by the Zeeman pairs (Figure 12). 
Pairs with negative field polarity have velocities ranging from 
$-$29.04 to $-$18.04 km s$^{-1}$ while those with positive 
field polarity have velocities ranging from $-$22.08 to $-$18.74 km s$^{-1}$.
Not only do the velocities of the pairs not vary systematically with 
position, but they also trace gas that is moving at
$\simeq$30 km s$^{-1}$ with respect to LSR velocity of G5.89. 
Hence, we find no indication of rotation in this clump of
maser emission.

We find good agreement between the magnetic field structure we present in 
Figure 11 with that presented in \cite{Fish05}.  In the E-group, 
\cite{Fish05} identify 4 pairs at 1667 MHz; similar to 
our observations, all pairs have positive magnetic field polarity with a 
median field strength of 1.5 mG.  In G5.89-S, \cite{Fish05}
 identify 3 pairs that appear to be part of the same clump of masers we 
observe 
in the south of the region. The field direction is negative 
for all pairs in this region, with a median field strength of $-$1.5 mG;
both measurements are consistent with the observations presented in this
paper. \cite{Fish05} only observe masers with positive radial 
velocities and hence do not detect the  
masers we identify toward the central region of the UC HII region
or those that trace the reversal in magnetic field polarity in G5.89-S. 

Each of the 1612 MHz masers is seen in both RCP and LCP.  Following
the same procedure described for the 1667 MHz masers, we find that
the velocity splitting between the different polarizations is 
greater than the velocity centroiding error for only one of the 
masers (feature number 3).  The velocity splitting is 0.13 km s$^{-1}$,
which is only $\simeq$15\% of the linewidth.  The B-field suggested 
by the pair is $-$1.05 mG; however, we note that the Zeeman splitting theory
changes significantly when the velocity splitting between polarizations 
is much less than a linewidth, so this value should be considered
 very uncertain. 

\subsection{Comparison with Methanol and Water Masers}

Methanol \citep{K04} and water \citep{H96} masers have also been detected 
toward G5.89.  The methanol masers do not appear to be
associated with the main-line OH masers.  While
the 1667 MHz and 1612 MHz OH masers are nearly all
projected toward the UC HII region, the 
methanol masers are offset $\sim$10 
arcseconds N of the center of the UC HII region,
with radial velocities ranging between 
2.88 km s$^{-1}$ and 15.48 km s$^{-1}$.  The distribution 
of water masers is primarily elongated NE-SW.
Two H$_2$O features are located along the E edge of the 
UC HII region with similar radial velocities as the 
1667 MHz OH masers in the same region, suggesting a possible
association.  

\section{Discussion}

Previous authors have suggested that the main-line OH masers
trace a rapidly expanding bipolar outflow \citep{Z90,W93}.  
If they do trace an outflow, we 
would expect the spatial distribution and velocity structure of 
the masers to resemble that of the outflowing gas.  

The 1667 MHz OH masers in G5.89 S and G5.89 Center are elongated
primarily along a NE-SW axis (PA=82$^\circ$)
similar to the SiO outflow \citep{S04}, but are 
confined to an area $\simeq$3 times smaller than that 
covered by SiO emission.  
We do not find a continuous velocity gradient along this axis,
in contrast to the observations of \cite{Z90}.  Further, the 
outflow structure suggested by \cite{Z90}
 (northern outflow cone pointed toward Earth and southern 
cone pointing away) is of opposite orientation to 
the SiO outflow, where the peak in the red-shifted lobe 
is located NE of the blue-shifted emission.

The factors listed above do not rule out the possibility
that the OH masers trace a bipolar outflow.  The 
confinement of OH masers to the face of 
the UC HII region can be explained if the excitation 
conditions for OH masers only exist very close to 
the base of the outflow, as suggested by \cite{Z90},
 while differences between the SiO(5-4) outflow
and the outflowing gas traced by the OH masers could simply
be due to the molecules tracing different components of the 
gas.  However, we suggest that a simpler explanation
may be that the OH masers are confined to
the postshock region at the interface of the ionized 
gas shell and the ambient molecular material, similar to 
the models discussed in \cite{Elitzur78}. 

The ionized gas appears to be expanding supersonically 
into the surrounding medium with an
expansion velocity of 35 km s$^{-1}$ \citep{Ac98}.
In the idealized case of expansion of a neutral shell,
we would expect all features projected on the
ionized shell to be blueshifted with respect to the systemic
velocity of the UC HII region.  Redshifted features would not 
appear projected near the central regions because the UC HII 
region is optically thick at 18 cm.  Those features located at
or close to the edge of the UC HII region should
have radial velocities comparable to the systemic
velocity of the central source.

Both of the above predicted characteristics are 
observed toward G5.89.  All masers projected in the central 
region of G5.89 are blueshifted by 30-42 km s$^{-1}$
with respect to the assumed systemic velocity of 
10 km s$^{-1}$, consistent with the 
expansion velocity of the UC HII region, while maser
emission projected on the edges of the UC HII region
exhibits radial velocities close to the systemic
velocity.  

If the masers do trace an expanding UC HII region, we would 
expect to observe large proper motions between masers located 
along the edges of the UC HII region and very small proper 
motions for masers in the center of the UC HII region. At 
present, we only have relative proper motion measurements
for masers along the edges of G5.89-E and G5.89-S.  The relative 
motions of these features do suggest that the masers are 
moving away from each other at high speeds (20-30 km s$^{-1}$); 
however, such high expansion speeds are also consistent with a 
scenario in which the masers trace an outflow.  A more definitive 
test awaits the determination of the tangential velocities of masers 
projected on the center of G5.89.

The observed magnetic field structure may also be naturally
explained in the context of a neutral shell 
shocked by the expanding HII region.  The ambient 
molecular gas surrounding G5.89 has
a density of $\simeq$1.2$\times$10$^{5}$ cm$^{-3}$ 
\citep{Ces91}. The resulting Alfven
speed, v$\rm_A \simeq$ 2 km s$^{-1}$, is significantly
less than the expansion speed of the UC HII region
(35 km s$^{-1}$), indicating that there is significantly
more energy in expansion than in the B-field.  Hence,
the magnetic field will be swept up and expand along
with the ionized gas shell.  An expanding shell 
carrying a magnetic field will have polarity
changes on the scale of the shell due to its curvature.
For example, a magnetic field that is initially directed 
N in the plane of the sky with no component along the 
line of sight will develop a positive line-of-sight 
component above the center of expansion and a negative 
line-of-sight component below the expansion center after being
swept up in the expanding ionized shell. This effect 
may explain the change in polarity observed between
the southernmost masers in G5.89 South and those located
several arcseconds N in G5.89 East and G5.89 Center.  An 
illustration of the effect is presented in \cite{Bou01}. 
The small-scale magnetic field reversal observed in G5.89 
South (Figure 10) may arise if the expanding shell crosses the 
path of a clump of gas in the ambient medium.  In this case 
a field reversal would occur on the clump size-scale.  
This hypothesis is supported by the presence of a foreground 
molecular cloud to the southwest of the UC HII region traced in 
1.3 mm continuum by \cite{Feldt99} and in 350 $\mu$m continuum 
by \cite{Mueller02}.

\section{Conclusions}

The large number ($>$50) and velocity range ($\simeq$50 km s$^{-1}$)
 of 1667 MHz OH masers toward G5.89 distinguishes this region  
from other massive star formation regions with OH masers. 
While it has been previously suggested that the OH masers
trace a rapidly expanding bipolar outflow, we suggest
that the maser emission may alternatively arise in the 
dense postshock region at the interface of the supersonically 
expanding UC HII region and the ambient molecular cloud.  

We identify 23 Zeeman pairs which indicate the
presence of $\lsim$ 2 mG magnetic fields.
An ordered reversal in magnetic
field direction  is present in the S region of G5.89.  
While the geometry hints at a toroidal pattern indicative
of a dense disk-like structure, the distribution and
velocity field of the OH masers suggests that this configuration 
is likely not the case.  The field reversal could alternatively 
be explained if the expanding shell of ionized 
gas sweeps up the magnetic field lines and crosses the path
of a neutral clump, thereby imprinting a field reversal 
on the clump size-scale.  

\acknowledgements{We thank the referee for helpful comments that 
improved the quality of the paper.  D.P.S. thanks Debra Shepherd 
and Christer Watson for useful discussions.  
{\it Facility: \facility{VLBA}}
}

\bibliography{journals_apj,mybib}

\clearpage

\begin{deluxetable}{ccccccccccc}
\tablenum{1}
\tabletypesize{\fontsize{7}{10}\selectfont}
\tablecaption{Summary of Observations\label{obs-table1}}
\tablehead{
\colhead{OH} & \colhead{RA} & \colhead{Dec} & \colhead{Bandwidth} &
\colhead{Spec. Res.} & \colhead{Beam Size} & \colhead{Beam PA} & 
\colhead{Noise} \\
\colhead{Line} &  \colhead{(2000)} & \colhead{(2000)}  & \colhead{(MHz)} &  
\colhead{(km s$^{-1}$)} & \colhead{(mas)} & \colhead{($^\circ$)}  & 
\colhead{(mJy beam$^{-1}$)}}
\startdata
1667    &  18:00:30.3820 &  $-$24:04:00.825 & 1.0 &  0.22  & 45 $\times$ 15 & 27 & 49   \\
1612    &  18:00:30.3820 &  $-$24:04:00.825 & 1.0 &  0.21  & 20 $\times$ 7  & 7  & 129  \\

\enddata
\end{deluxetable}

\begin{deluxetable}{crrccc}
\tablenum{2}
\tabletypesize{\fontsize{7}{10}\selectfont}
\tablecaption{G5.886$-$0.393 1667 MHz LCP Maser Parameters\label{1667L}}
\tablehead{
  \colhead{Maser} &
  \colhead{$\Delta \alpha$} &
  \colhead{$\Delta \delta$} &
  \colhead{Vel.} &
  \colhead{$\Delta v$} &
  \colhead{Flux Density} \\
  \colhead{Feature} &
  \colhead{(mas)} &
  \colhead{(mas)} &
  \colhead{(km s$^{-1}$)} &
  \colhead{(km s$^{-1}$)} &
  \colhead{(Jy)}}
\startdata
 1   &     $-$628   &    $-$2857   &  15.29   &  0.30   &  0.66 \\
 2   &     $-$566   &    $-$2966   &  14.51   &  0.39   &  0.51 \\
 3   &     $-$549   &    $-$2733   &  14.68   &  0.50   &  0.55 \\
 4   &     $-$519   &    $-$2803   &  14.09   &  0.85   &  0.70 \\
 5   &     $-$517   &    $-$2867   &  15.30   &  0.40   &  0.47 \\
 6   &     $-$514   &    $-$2648   &  14.02   &  0.45   &  1.34 \\
 7   &     $-$513   &    $-$2706   &  14.50   &  0.29   &  2.05 \\
 8   &     $-$492   &    $-$2745   &  13.78   &  0.41   &  0.91 \\
 9   &     $-$428   &    $-$2677   &  11.38   &  0.55   &  0.64 \\
10   &     $-$421   &    $-$2517   &  13.36   &  0.38   &  4.62 \\
11   &     $-$399   &    $-$2491   &  13.15   &  0.96   &  0.60 \\
12   &     $-$382   &    $-$1862   & $-$20.41   &  0.48   &  0.79 \\
13   &     $-$371   &     $-$232   & $-$28.17   &  0.54   &  0.49 \\
14   &     $-$365   &    $-$1659   & $-$18.20   &  0.63   &  1.99 \\
15   &     $-$339   &    $-$2687   &  15.17   &  0.57   &  0.54 \\
16   &     $-$287   &    $-$1702   & $-$18.83   &  0.50   &  7.20 \\
17   &     $-$285   &    $-$1775   & $-$21.08   &  0.42   &  2.85 \\
18   &     $-$285   &        1     & $-$31.53   &  0.47   &  0.91 \\
19   &     $-$273   &    $-$1734   & $-$20.41   &  0.33   &  4.22 \\
20   &     $-$225   &    $-$1802   & $-$22.19   &  0.48   &  2.50 \\
21   &     $-$183   &       13     & $-$29.33   &  0.43   &  2.09 \\
22   &     $-$182   &    $-$2006   & $-$28.86   &  0.77   &  0.92 \\
23   &     $-$180   &     $-$425   & $-$22.03   &  0.59   &  0.55 \\
24   &     $-$162   &     $-$440   & $-$21.37   &  0.55   &  0.63 \\
25   &     $-$139   &    $-$1991   & $-$26.98   &  1.14   &  0.69 \\
26   &     $-$122   &    $-$1820   & $-$20.80   &  0.62   &  1.09 \\
27   &     $-$121   &    $-$1990   & $-$26.37   &  0.57   &  2.80 \\
28   &     $-$111   &     $-$131   & $-$29.66   &  0.36   &  2.16 \\
29   &      $-$96   &    $-$1810   & $-$20.13   &  0.47   &  0.66 \\
30   &      $-$86   &     $-$102   & $-$29.14   &  0.53   &  1.55 \\
31   &      $-$38   &    $-$1812   & $-$19.22   &  0.45   &  9.80 \\
32   &      $-$34   &     $-$267   & $-$25.56   &  0.35   &  3.77 \\
33   &       $-$5   &    $-$1790   & $-$17.92   &  0.51   &  3.40 \\
34   &      118   &     $-$287     & $-$24.67   &  0.27   &  1.56 \\
35   &      232   &     $-$261     & $-$26.82   &  0.42   &  1.11 \\
36   &      260   &     $-$292     & $-$25.91   &  0.27   &  0.77 \\
37   &      291   &     $-$313     & $-$25.13   &  0.35   &  4.30 \\
38   &      318   &     $-$293     & $-$25.96   &  0.33   &  1.42 \\
39   &      329   &     $-$430     & $-$23.15   &  0.48   &  1.86 \\
40   &      362   &    $-$1481     & $-$18.32   &  0.46   &  0.68 \\
41   &      378   &     $-$314     & $-$26.82   &  0.34   &  2.93 \\
42   &      395   &     $-$350     & $-$26.11   &  0.56   &  2.71 \\
43   &      476   &    $-$1303     & $-$24.49   &  0.41   &  0.60 \\
44   &      497   &     $-$569     & $-$23.00   &  0.57   &  0.58 \\
45   &      566   &      259       & $-$31.17   &  0.78   &  1.78 \\
46   &      616   &      282       & $-$32.13   &  0.38   &  0.78 \\
47   &      626   &    $-$1188     & $-$24.37   &  0.34   &  0.71 \\
48   &      657   &    $-$1156     & $-$23.13   &  0.46   &  0.82 \\
49   &     1016   &      343       & $-$26.95   &  0.67   &  1.69 \\
50   &     1017   &      214       & $-$24.12   &  0.33   &  0.61 \\
51   &     1028   &      311       & $-$26.29   &  0.38   &  2.36 \\
52   &     2128   &     $-$778     & $-$29.88   &  0.34   &  1.01 \\
53   &     2221   &     $-$183     & $-$11.74   &  0.36   &  0.63 \\
54   &     2756   &     $-$402     & $-$15.05   &  0.33   &  2.93 \\
55   &     2829   &     $-$425     & $-$16.67   &  0.45   &  7.50 \\
56   &     3342   &    $-$1980   &  11.02   &  0.32   &  0.81 \\
57   &     3369   &     $-$768   &   8.33   &  0.34   &  0.34 \\
58   &     3410   &     $-$760   &   8.92   &  0.32   &  0.44 \\
59   &     3470   &     $-$686   &   9.78   &  0.27   &  6.22 \\

\enddata
\end{deluxetable}
\clearpage

\begin{deluxetable}{crrccc}
\tablenum{3}
\tabletypesize{\fontsize{7}{10}\selectfont}
\tablecaption{G5.886$-$0.393 1667 MHz RCP Maser Parameters\label{1667R}}
\tablehead{
  \colhead{Maser} &
  \colhead{$\Delta \alpha$} &
  \colhead{$\Delta \delta$} &
  \colhead{Vel.} &
  \colhead{$\Delta v$} &
  \colhead{Flux Density} \\
  \colhead{Feature} &
  \colhead{(mas)} &
  \colhead{(mas)} &
  \colhead{(km s$^{-1}$)} &
  \colhead{(km s$^{-1}$)} &
  \colhead{(Jy)}}
\startdata
 1   &     $-$626   &    $-$2858   &  15.19   &  0.34   &  1.52 \\
 2   &     $-$530   &    $-$2813   &  14.29   &  0.23   &  0.81 \\
 3   &     $-$516   &    $-$2801   &  13.74   &  0.58   &  1.34 \\
 4   &     $-$495   &    $-$2668   &  14.27   &  0.31   &  0.44 \\
 5   &     $-$485   &    $-$2699   &  14.59   &  0.32   &  0.56 \\
 6   &     $-$474   &    $-$2585   &  11.52   &  0.43   &  0.30 \\
 7   &     $-$442   &    $-$2795   &  14.51   &  0.28   &  0.24 \\
 8   &     $-$428   &    $-$2678   &  11.37   &  0.59   &  0.38 \\
 9   &     $-$421   &     $-$309   & $-$26.76   &  0.39   &  0.87 \\
10   &     $-$420   &    $-$2468   &  11.87   &  0.48   &  0.47 \\
11   &     $-$415   &    $-$2510   &  12.66   &  0.38   &  0.50 \\
12   &     $-$389   &    $-$2768   &   7.96   &  0.42   &  0.21 \\
13   &     $-$380   &    $-$1863   & $-$19.72   &  0.44   &  0.99 \\
14   &     $-$366   &     $-$228   & $-$28.10   &  0.57   &  1.20 \\
15   &     $-$356   &    $-$2695   &  15.20   &  0.26   &  0.42 \\
16   &     $-$339   &    $-$2546   &   7.04   &  0.68   &  0.46 \\
17   &     $-$299   &    $-$2641   &   7.05   &  0.43   &  0.64 \\
18   &     $-$285   &       $-$7   & $-$31.61   &  0.40   &  1.18 \\
19   &     $-$275   &    $-$1695   & $-$18.65   &  0.65   &  2.79 \\
20   &     $-$270   &    $-$1737   & $-$19.95   &  0.33   &  1.53 \\
21   &     $-$269   &     $-$197   & $-$28.50   &  0.25   &  0.63 \\
22   &     $-$242   &     $-$328   & $-$25.15   &  0.35   &  0.92 \\
23   &     $-$218   &    $-$1799   & $-$21.97   &  0.47   &  0.98 \\
24   &     $-$214   &       $-$5   & $-$29.72   &  0.35   &  1.49 \\
25   &     $-$184   &    $-$2005   & $-$29.23   &  0.67   &  1.09 \\
26   &     $-$184   &       16     & $-$29.34   &  0.35   &  2.03 \\
27   &     $-$141   &    $-$1996   & $-$27.53   &  0.94   &  0.72 \\
28   &     $-$126   &     $-$132   & $-$29.64   &  0.29   &  1.25 \\
29   &     $-$122   &    $-$1987   & $-$26.86   &  0.78   &  1.15 \\
30   &     $-$120   &    $-$1819   & $-$21.04   &  0.62   &  1.27 \\
31   &     $-$100   &     $-$260   & $-$26.56   &  0.43   &  0.90 \\
32   &      $-$95   &    $-$1808   & $-$20.38   &  0.47   &  1.12 \\
33   &      $-$92   &     $-$121   & $-$29.57   &  1.24   &  0.64 \\
34   &      $-$86   &     $-$103   & $-$29.16   &  0.51   &  1.05 \\
35   &      $-$62   &     $-$153   & $-$30.27   &  0.30   &  0.47 \\
36   &      $-$32   &     $-$266   & $-$25.33   &  0.49   &  2.41 \\
37   &      $-$16   &     $-$239   & $-$28.49   &  0.28   &  0.81 \\
38   &       $-$5   &    $-$1781   & $-$18.16   &  0.40   &  2.73 \\
39   &       55   &     $-$236     & $-$26.87   &  0.29   &  0.51 \\
40   &      223   &     $-$444     & $-$20.60   &  0.37   &  0.68 \\
41   &      235   &     $-$270     & $-$26.54   &  0.36   &  0.67 \\
42   &      286   &     $-$309     & $-$25.10   &  0.30   &  0.64 \\
43   &      333   &     $-$430     & $-$22.86   &  0.42   &  1.70 \\
44   &      380   &     $-$310     & $-$26.61   &  0.31   &  0.70 \\
45   &      476   &     $-$1301    & $-$23.97   &  0.39   &  0.55 \\
46   &      498   &     $-$572     & $-$22.52   &  0.46   &  0.65 \\
47   &      568   &      266       & $-$31.48   &  0.79   &  2.61 \\
48   &     1018   &      339       & $-$26.84   &  0.50   &  3.31 \\
49   &     1039   &      308       & $-$26.14   &  0.36   &  0.90 \\
50   &     2111   &     $-$779     & $-$30.09   &  0.30   &  0.77 \\
51   &     2253   &     $-$201     & $-$11.54   &  0.38   &  0.99 \\
52   &     2758   &     $-$401     & $-$14.86   &  0.31   &  1.73 \\
53   &     2821   &     $-$276     & $-$16.35   &  0.42   &  0.65 \\
54   &     2825   &     $-$426     & $-$16.20   &  0.49   &  2.65 \\
55   &     3473   &     $-$685     &  10.28   &  0.22   &  4.88 \\
\enddata
\end{deluxetable}
\clearpage

\begin{deluxetable}{crrccc}
\tablenum{4}
\tabletypesize{\fontsize{7}{10}\selectfont}
\tablecaption{G5.886$-$0.393 1612 MHz Maser Parameters\label{1612}}
\tablehead{
  \colhead{Maser} &
  \colhead{$\Delta \alpha$} &
  \colhead{$\Delta \delta$} &
  \colhead{Vel.} &
  \colhead{$\Delta v$} &
  \colhead{Flux Density} \\
  \colhead{Feature} &
  \colhead{(mas)} &
  \colhead{(mas)} &
  \colhead{(km s$^{-1}$)} &
  \colhead{(km s$^{-1}$)} &
  \colhead{(Jy)}}
\startdata
\multicolumn{6}{l}{1612 MHz LCP}\\*
 1   &     $-$181   &      109   & $-$19.26   &  0.65   &  3.30 \\
 2   &     $-$157   &      163   & $-$20.51   &  0.87   &  2.26 \\
 3   &        0   &        0     & $-$21.31   &  0.85   & 12.05 \\
\multicolumn{6}{l}{1612 MHz RCP}\\*		   
 1   &     $-$179   &      106   & $-$19.37   &  0.84   &  4.63 \\
 2   &     $-$157   &      159   & $-$20.43   &  0.80   &  2.73 \\
 3   &        0   &        0     & $-$21.43   &  1.00   & 12.39 \\

\enddata
\end{deluxetable}
\clearpage

\begin{deluxetable}{rrcrrccrr}
 \tablenum{5}
 \tablewidth{410pt}
 \tabletypesize{\fontsize{7}{10}\selectfont}
 \tablecaption{Zeeman Pairs\label{zeeman}}
 \tablehead{
  \multicolumn{3}{c}{\hrulefill RCP \hrulefill}
  &\multicolumn{3}{c}{\hrulefill LCP \hrulefill}
 &&\multicolumn{2}{c}{Separation}\\
  \colhead{$\Delta\alpha$} &
  \colhead{$\Delta\delta$} &
  \colhead{Velocity} &
  \colhead{$\Delta\alpha$} &
  \colhead{$\Delta\delta$} &
  \colhead{Velocity} &
  \colhead{$B$} &
  \colhead{Ang.} &
  \colhead{Lin.} \\
  \colhead{(mas)} &
  \colhead{(mas)} &
  \colhead{(km s$^{-1}$)} &
  \colhead{(mas)} &
  \colhead{(mas)} &
  \colhead{(km s$^{-1}$)} &
  \colhead{(mG)} &
  \colhead{(mas)} &
  \colhead{(AU)} \\ 
 }
\startdata
$-$516  &    $-$2801  &    13.74    &  $-$519  & $-$2803  &    14.09    &    $-$0.96  &   4.16  &   8.32 \\
$-$415  &    $-$2510  &    12.66    &  $-$421  & $-$2517  &    13.36    &    $-$1.97  &   8.53  &  17.07 \\
$-$380  &    $-$1863  &   $-$19.72  &  $-$382  & $-$1862  &   $-$20.41  &     1.94  &   1.82  &   3.65 \\
$-$276  &    $-$1695  &   $-$18.65  &  $-$287  & $-$1702  &   $-$18.83  &     0.51  &  13.62  &  27.25 \\
$-$270  &    $-$1737  &   $-$19.95  &  $-$273  & $-$1734  &   $-$20.41  &     1.29  &   4.39  &   8.78 \\
$-$218  &    $-$1799  &   $-$21.97  &  $-$225  & $-$1802  &   $-$22.19  &     0.62  &   7.46  &  14.93 \\
$-$184  &    $-$2005  &   $-$29.23  &  $-$182  & $-$2006  &   $-$28.86  &    $-$1.05  &   1.85  &   3.70 \\
$-$141  &    $-$1996  &   $-$27.53  &  $-$139  & $-$1991  &   $-$26.98  &    $-$1.57  &   5.37  &  10.73 \\
$-$122  &    $-$1987  &   $-$26.86  &  $-$121  & $-$1990  &   $-$26.37  &    $-$1.37  &   2.74  &   5.48 \\
$-$120  &    $-$1819  &   $-$21.04  &  $-$122  & $-$1820  &   $-$20.80  &    $-$0.67  &   2.11  &   4.23 \\
 $-$32  &     $-$266  &   $-$25.33  &   $-$34  &  $-$267  &   $-$25.56  &     0.65  &   2.44  &   4.88 \\
  $-$5  &    $-$1781  &   $-$18.16  &    $-$5  & $-$1790  &   $-$17.92  &    $-$0.68  &   8.98  &  17.96 \\
 235  &     $-$270  &   $-$26.54  &   232      &  $-$261      &   $-$26.82  &     0.81  &   8.84  &  17.68 \\
 333  &     $-$430  &   $-$22.86  &   329  &  $-$430     &   $-$23.15  &     0.83  &   3.57  &   7.14 \\
 380  &     $-$310  &   $-$26.61  &   378  &  $-$314      &   $-$26.82  &     0.59  &   4.45  &   8.89 \\
 476  &    $-$1301  &   $-$23.97  &   476  &  $-$1303     &   $-$24.49  &     1.49  &   2.25  &   4.50 \\
 498  &     $-$572  &   $-$22.52  &   497  &  $-$569      &   $-$23.00  &     1.35  &   3.20  &   6.40 \\
 568  &      266    &   $-$31.48  &   566  &   259        &   $-$31.17  &    $-$0.88  &   7.03  &  14.06 \\
1018  &      339    &   $-$26.84  &  1016  &   343        &   $-$26.95  &     0.29  &   5.14  &  10.28 \\
1039  &      308    &   $-$26.14  &  1028  &   311        &   $-$26.29  &     0.45  &  10.57  &  21.14 \\
2758  &     $-$401  &   $-$14.86  &  2756  &  $-$402      &   $-$15.05  &     0.52  &   1.86  &   3.72 \\
2825  &     $-$426  &   $-$16.20  &  2829  &  $-$425      &   $-$16.67  &     1.34  &   4.31  &   8.63 \\
3473  &     $-$685  &    10.28    &  3470  &    $-$686      &     9.78  &     1.41  &   3.14  &   6.28 \\

\enddata
\end{deluxetable}

\clearpage



\begin{figure}
\figurenum{1}
\epsscale{.85}
\plotone{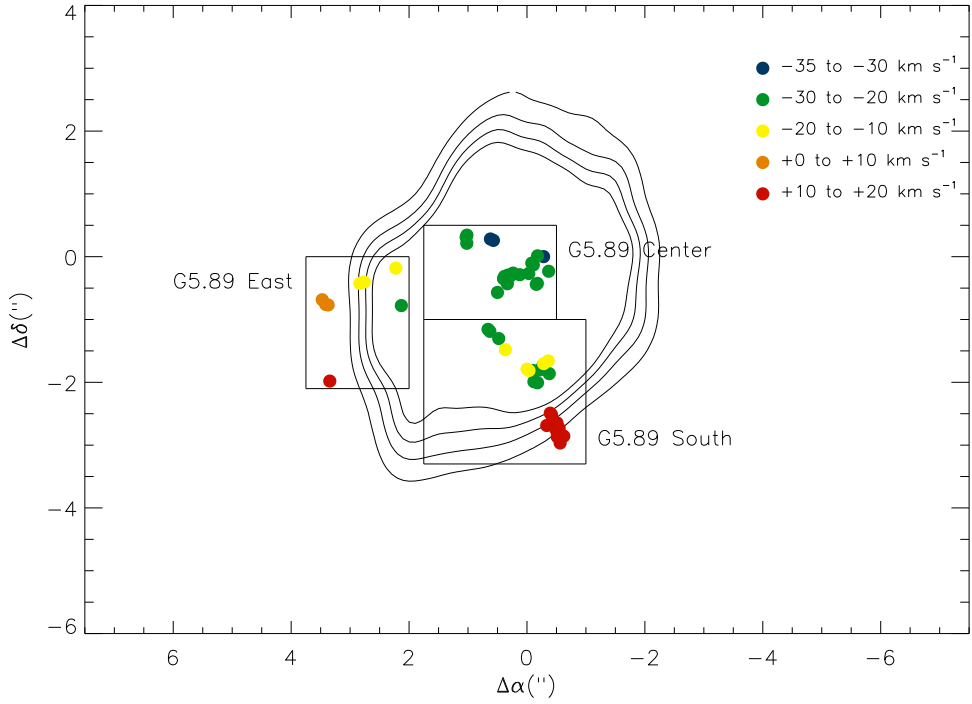}
\caption{Left Circular Polarized 1667 MHz OH masers overlaid on 8.5 GHz 
continuum.   The origin corresponds to $\alpha_{2000}$=18$^h$00$^m$30\fs382 
$\delta_{2000}$=$-$24$^\circ$04'00\farcs829.  The 8.5 GHz continuum 
data was obtained with the VLA in AB configuration (Churchwell 2003, 
private communication) and has a beam size of 0.6$\times$0.5 arcseconds.  
The contour levels of the radio continuum image are 0.03, 0.06, 0.10, 
0.13 Jy beam$^{-1}$.
}
\label{L67}
\end{figure}

\begin{figure}
\figurenum{2}
\epsscale{.85}
\plotone{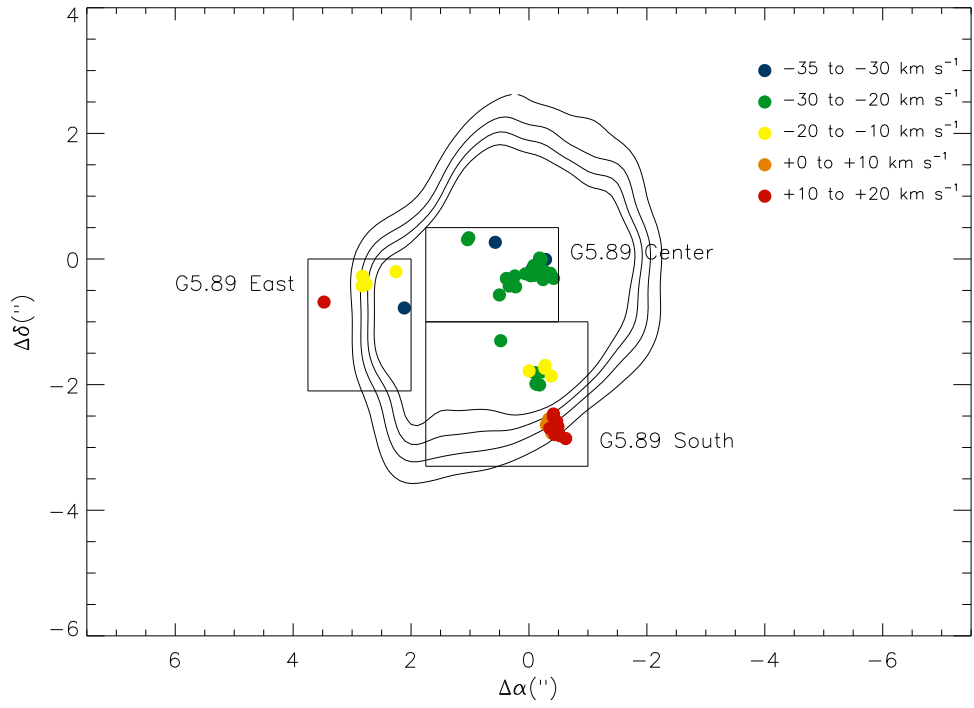}
\caption{Right Circular Polarized 1667 MHz OH masers overlaid on 8.5 GHz continuum. 
See Figure 1 for details.
}
\label{R67}
\end{figure}

\begin{figure}
\figurenum{3}
\epsscale{.85}
\plotone{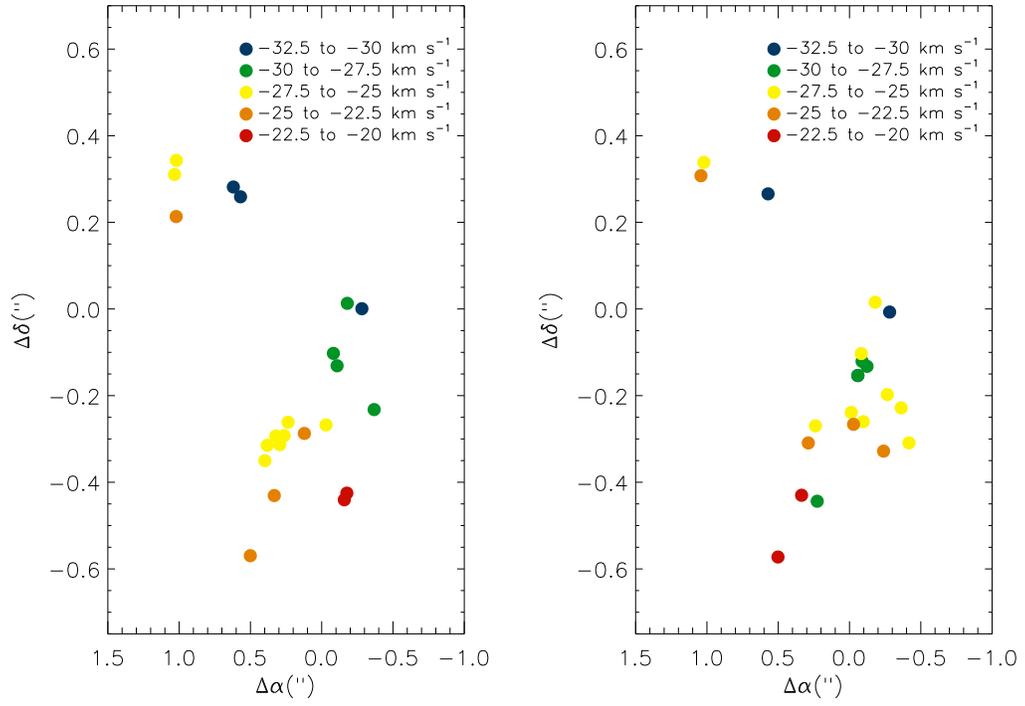}
\caption{Left: Positions of LCP 1667 MHz masers in G5.89 Center.
Right: Positions of RCP 1667 MHz masers in G5.89 Center. See 
Figure 1 for details.
}
\label{center}
\end{figure}

\begin{figure}
\figurenum{4}
\epsscale{.85}
\plotone{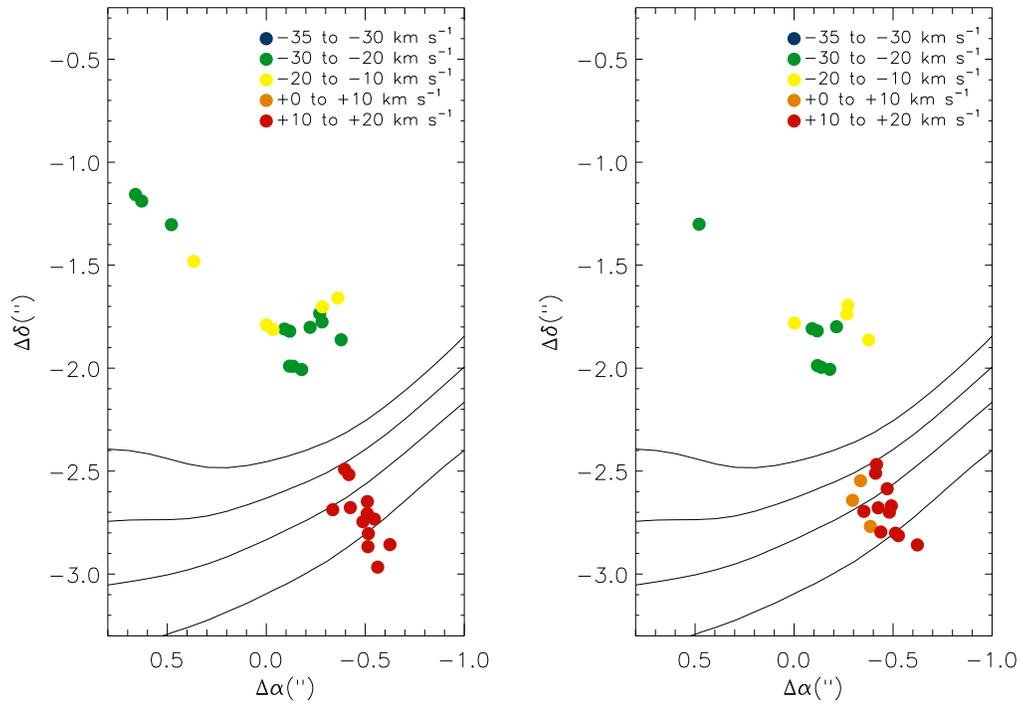}
\caption{Left: Positions of LCP 1667 MHz masers in G5.89 South.
Right: Positions of RCP 1667 MHz masers in G5.89 South. See 
Figure 1 for details.
}
\label{south}
\end{figure}

\begin{figure}
\figurenum{5}
\epsscale{.85}
\plotone{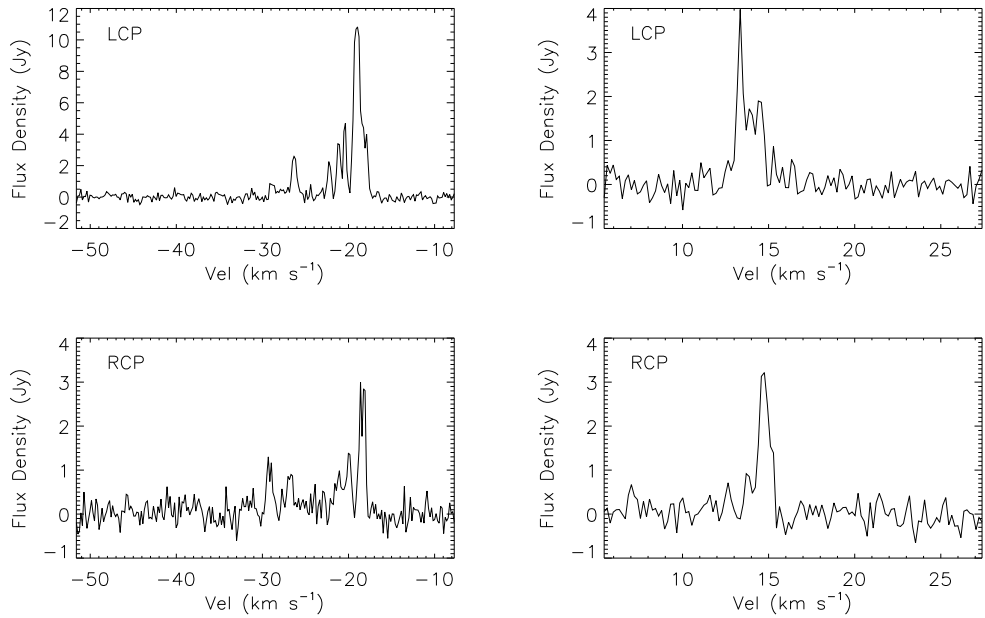}
\caption{Spectrum of 1667 MHz emission averaged over S region as specified 
in Figure 1. The LCP spectra is displayed in the top panel and the RCP 
spectrum is displayed in the bottom panel. The beam size is 45$\times$15 mas 
at 1667 MHz. Velocities are with respect to the local standard of rest.
Line emission from OH masers spans 45 km s$^{-1}$ in a region 1.8x1.4 arcseconds in size. 
}
\label{southspec}
\end{figure}

\begin{figure}
\figurenum{6}
\epsscale{.85}
\plotone{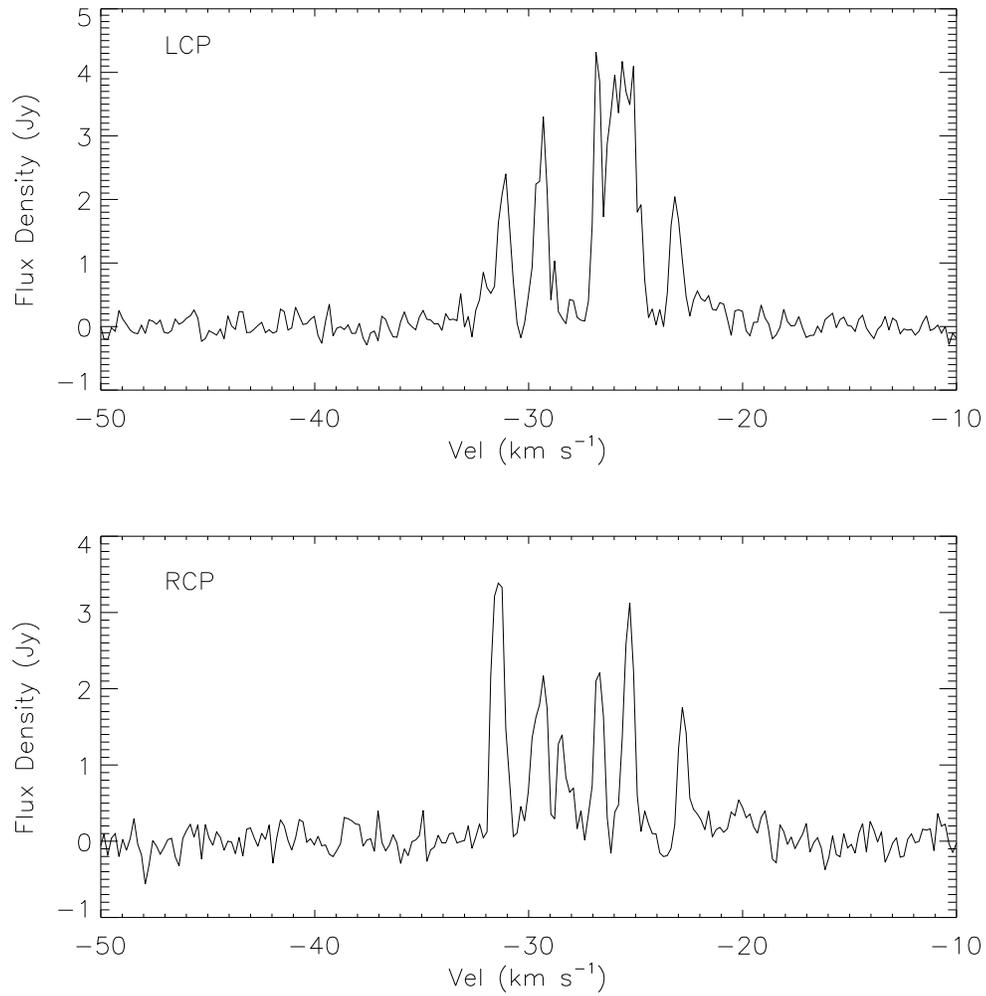}
\caption{Spectrum of 1667 MHz emission averaged over the central region 
of G5.89 (see Fig 1). The LCP spectra is displayed in the top panel and 
the RCP spectrum is displayed in the bottom panel. The beam size is 
45$\times$15 mas at 1667 MHz.  Masers are only 
detected in this region at negative LSR velocities.
}
\label{north}
\end{figure}

\begin{figure}
\figurenum{7}
\epsscale{.85}
\plotone{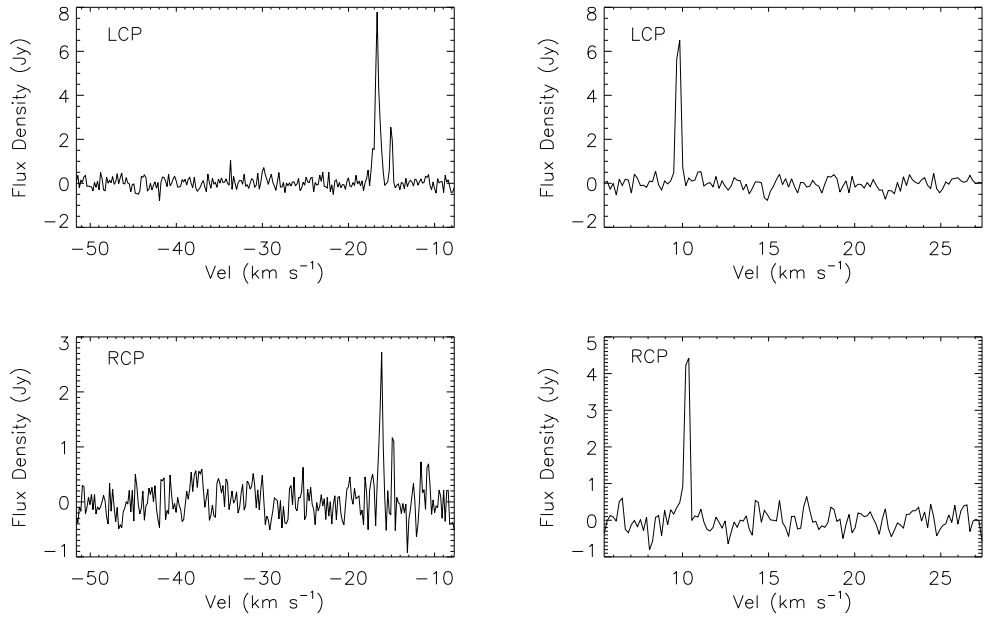}
\caption{Spectra of LCP 1667 MHz emission averaged over E region of G5.89
(see Fig. 1).  The LCP spectra is displayed in the top panel and 
the RCP spectrum is displayed in the bottom panel. The beam size is 
45$\times$15 mas at 1667 MHz.
}
\label{eastlcp}
\end{figure}

\begin{figure}
\figurenum{8}
\epsscale{.85}
\plotone{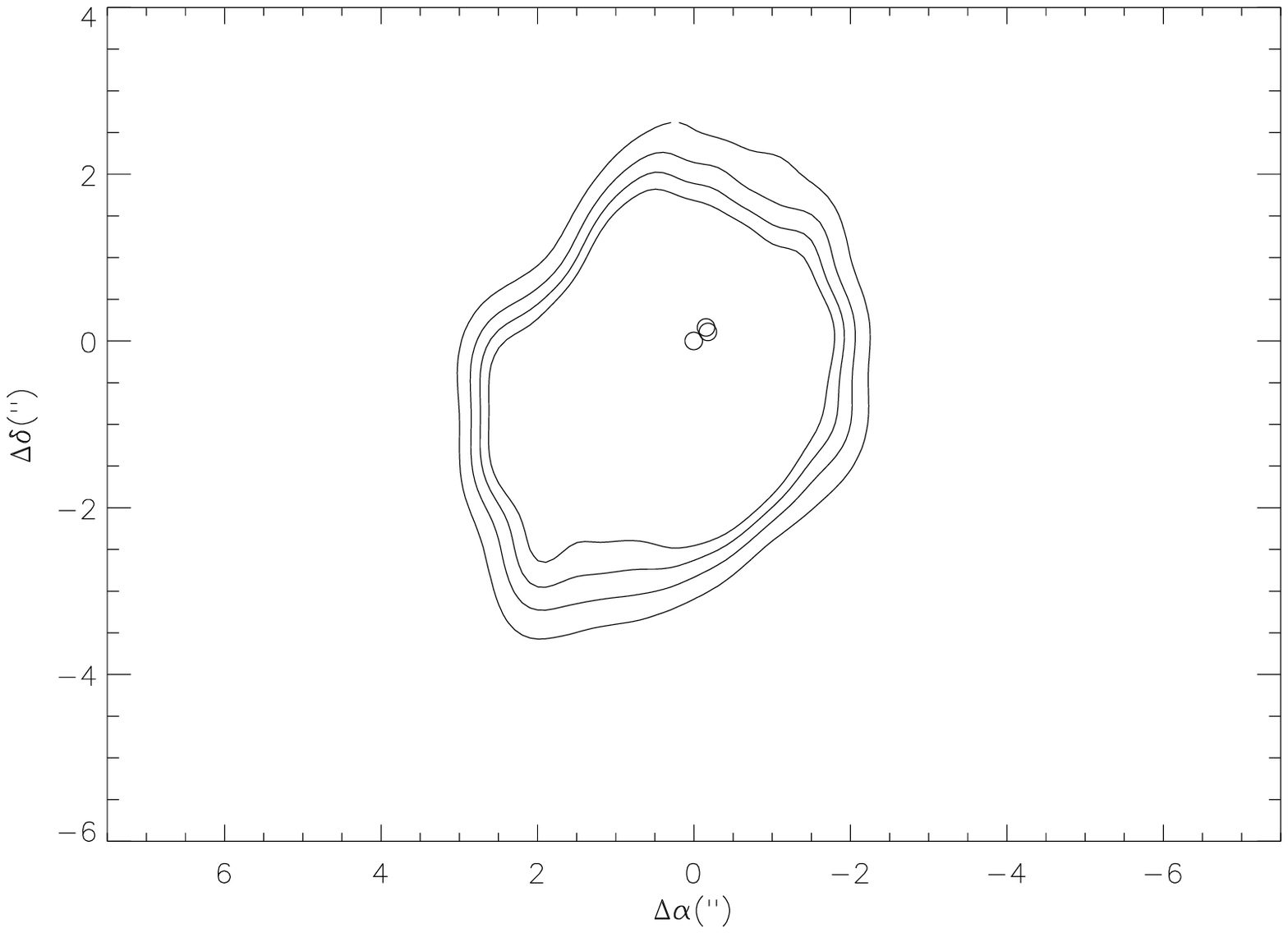}
\caption{Left Circular Polarized 1612 MHz OH masers overlaid on 8.5 GHz 
continuum. The beam size is 20$\times$7 mas at 1612 MHz.
}
\label{L12}
\end{figure}

\begin{figure}
\figurenum{9}
\epsscale{.85}
\plotone{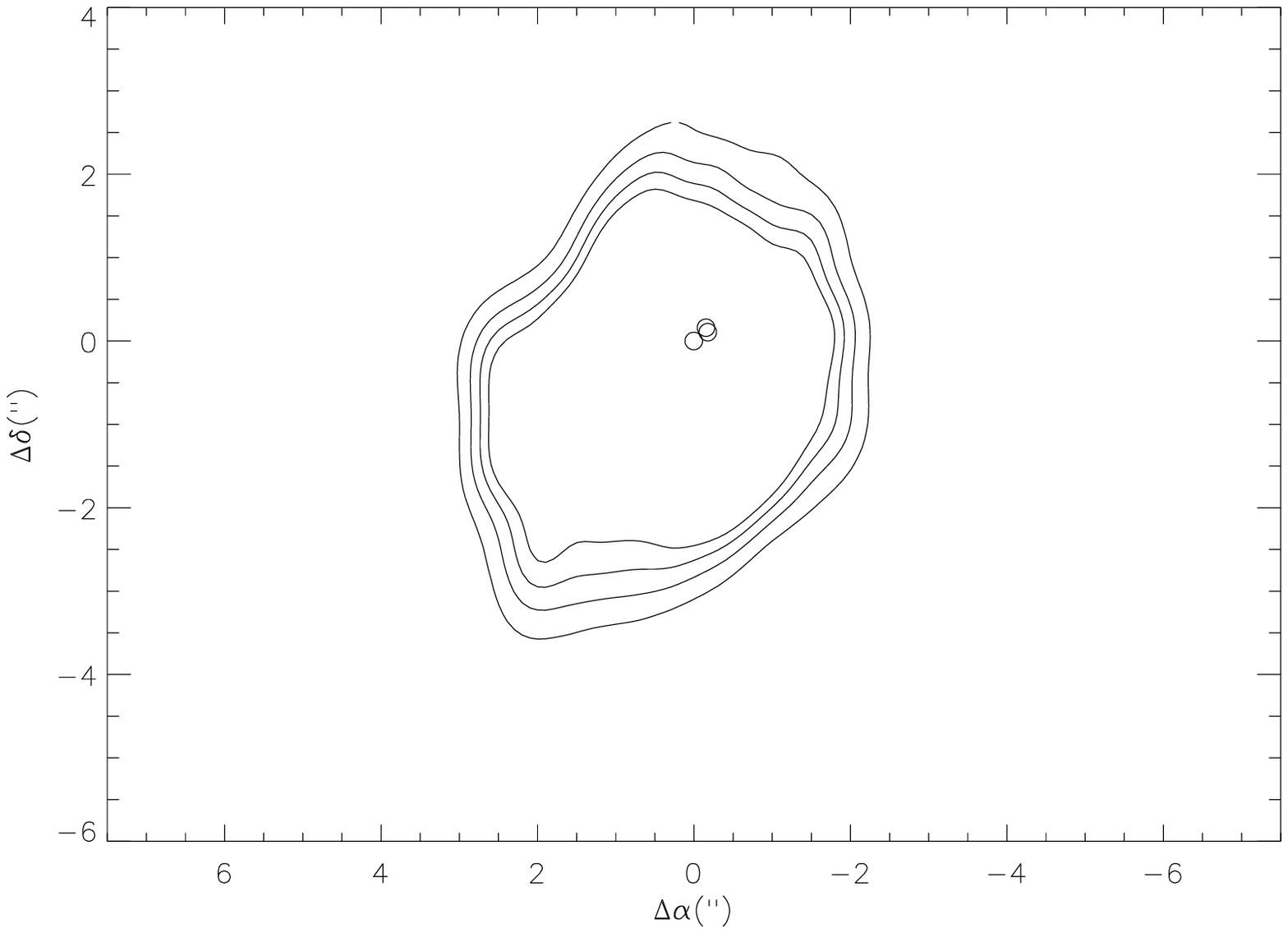}
\caption{Right Circular Polarized 1612 MHz OH masers overlaid on 8.5 GHz 
continuum. The beam size is 20$\times$7 mas at 1612 MHz.
}
\label{R12}
\end{figure}

\begin{figure}
\figurenum{10}
\epsscale{.85}
\plotone{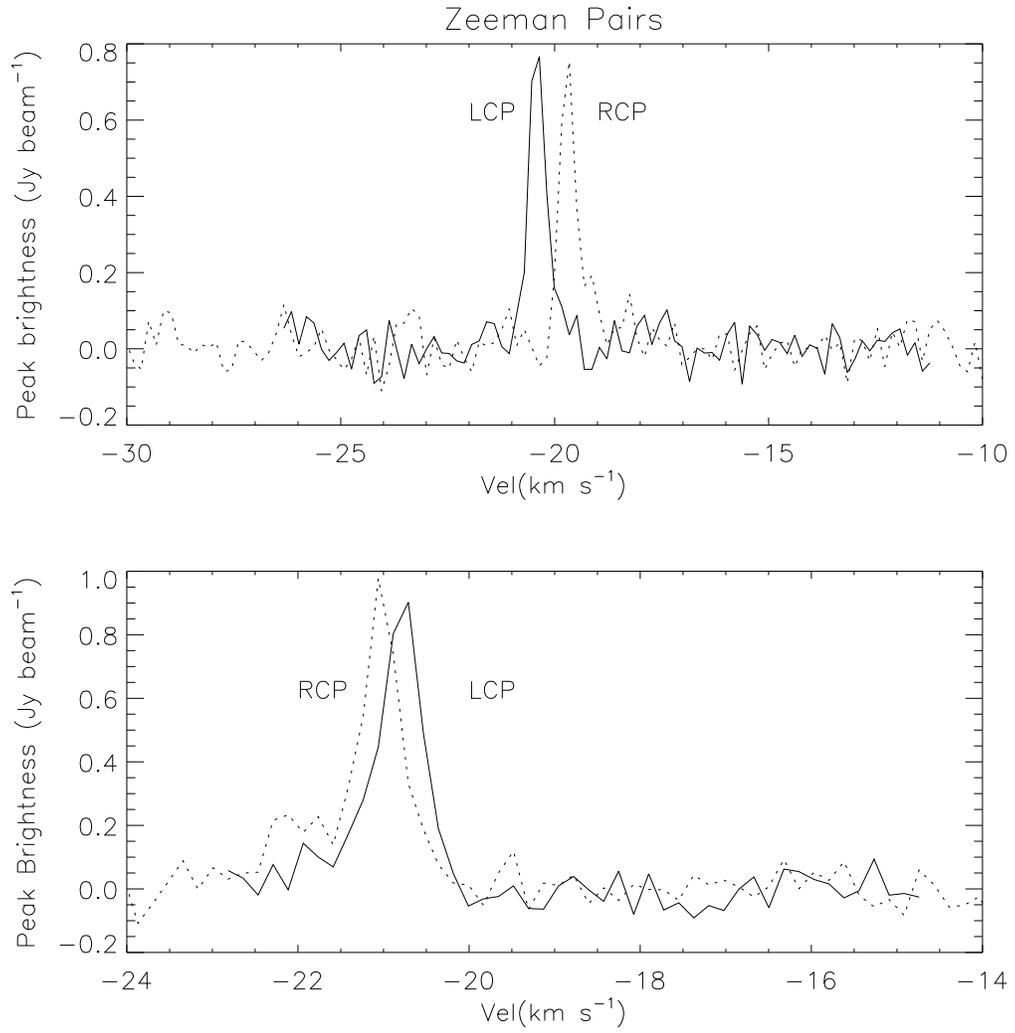}
\caption{VLBA spectra of two typical Zeeman pairs at 1667 MHz in 
both LCP and RCP.
}
\label{pol}
\end{figure}

\begin{figure}
\figurenum{11}
\epsscale{.85}
\plotone{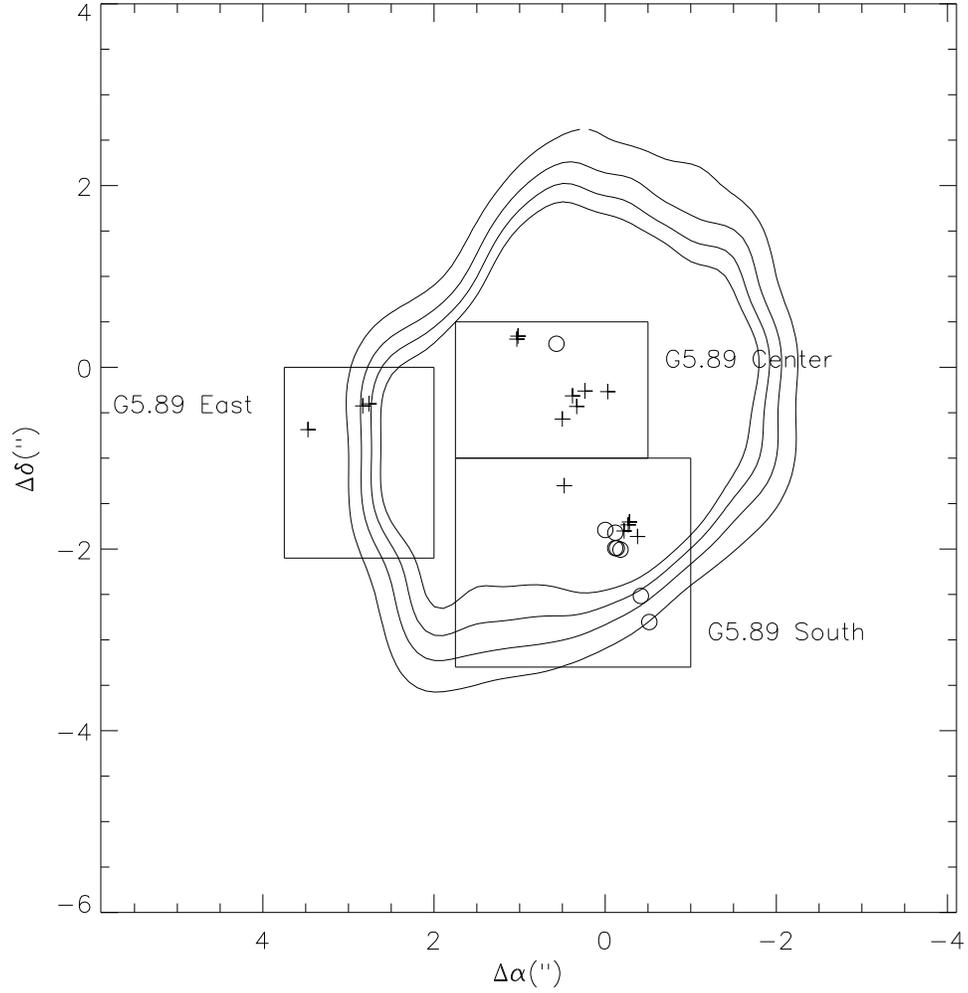}
\caption{Zeeman Pairs identified toward G5.89 at 1667 MHz overlaid 
over 8.5 GHz continuum emission.  Circles are representive of a negative 
magnetic field, whereas crosses indicate a positive B-field polarity.    
}
\label{bfield}
\end{figure}
\clearpage

\begin{figure}
\figurenum{12}
\epsscale{.85}
\plotone{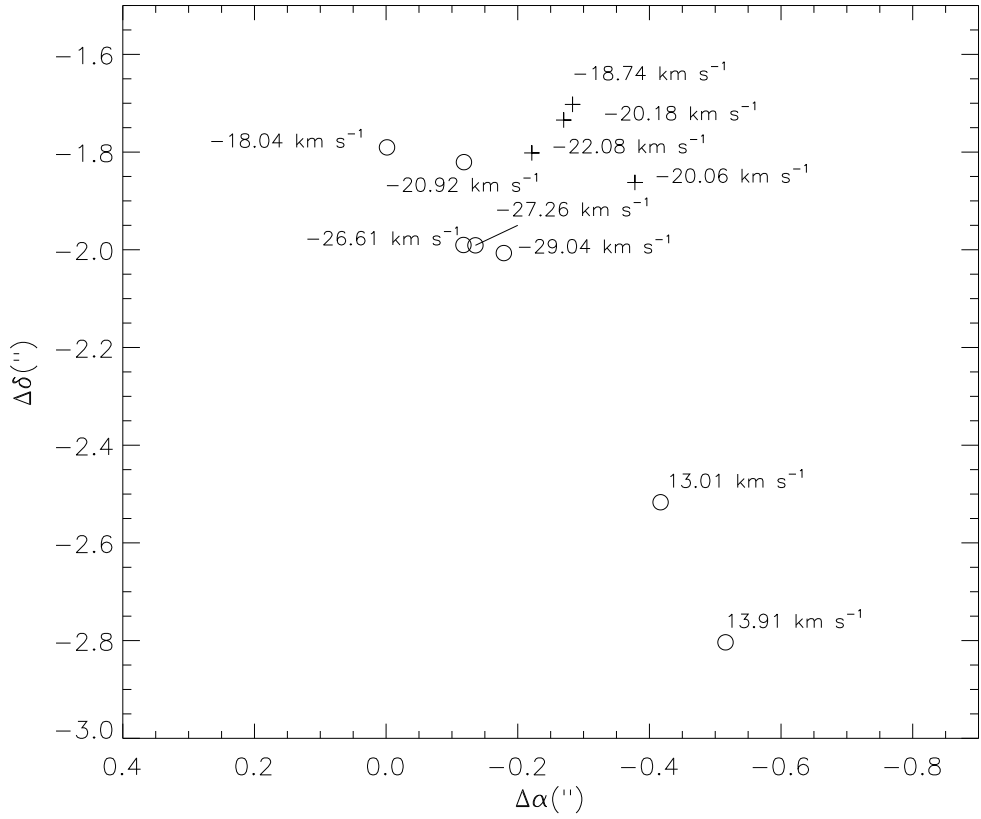}
\caption{Magnetic-field structure in G5.89 South as traced by 1667 MHz OH 
masers determined using the VLBA of the NRAO.  Circles are representative 
of a negative magnetic field, whereas crosses indicate a positive B-field 
polarity. Radial velocities (after correction for the Zeeman effect) are 
noted next to each feature.  The direction 
of the magnetic field changes over $\simeq$0.3 arcseconds, corresponding to 
600 AU at the distance of G5.89. 
}
\label{bzoom}
\end{figure}
\clearpage

\end{document}